\documentstyle[epsf,twoside,fleqn,espcrc2]{article}

\newcommand{\AmS}{{\protect\the\textfont2
   A\kern-.1667em\lower.5ex\hbox{M}\kern-.125emS}}
\def\z0{\rm Z^0}
\newcommand{\as}{\alpha_{\rm s}}

\newcommand{\oaaa}{{\cal O}(\as^3)}
\newcommand{\epem}{\rm e^+\rm e^-}

\newcommand{\amz}{\as(M_{\rm Z})}

\def\wamz{\overline{\as}(M_{\rm Z})}
\def\dwas{\Delta\overline{\as}}
\def\mz{M_{\rm Z^0}}

\def\d2{D_2}
\def\oq{\char'134}
\def\lamsb{\Lambda_{\overline{MS}}}

\def\m2{\mu^2}
\def\q{\rm q}

\def\p{\rm p}

\def\q2{Q^2}
\def\asq{\as (\q2 )}

\title{$\as$ 2002}

\author{S. Bethke\address{Max Planck Institut f\"ur Physik \\
         80805 M\"unchen, Germany}%
        }

\begin{document}

\begin{abstract}
An update of measurements of the strong coupling constant
$\as$ is given, representing the status of
September 2002.
The results convincingly prove the energy dependence of $\as$ and are in 
excellent agreement with the expectations of Quantum Chromodynamics, QCD.
Evolving all results to the rest energy of the $\z0$ boson, 
the new world average of 
$\amz$ is determined from measurements which are based on QCD calculations in 
complete NNLO 
perturbation theory, giving
$$ \amz = 0.1183 \pm 0.0027 \ . $$

%
% following 4 lines for preprint only:
\vskip-95mm
   {\small \noindent
  Talk presented at the {\it QCD 02 High-Energy Physics
    International Conference
    in Quantum Chromodynamics}, Montpellier (France)
    July 2-9, 2002.} 
   \begin{flushright} {\large MPI-PhE/2002-17} \\
    {\large November 2002}\\
%    {revised: November 1997}
  \end{flushright}
\vskip75mm 

\end{abstract}

\maketitle

\section{INTRODUCTION}

The coupling constant of the Strong Interactions, $\as$, is 
one of the most fundamental parameters of nature which must be determined by 
experiment.
Many new results, experimental studies as well as improved theoretical 
calculations, continue to be provided each year, such that regular updates 
of summaries of measurements of $\as$ are mandatory.

In this contribution, an update of 
the most recent results on determinations of $\as$
which were published since summer 2000 is given.
It is organised as an incremental addition to a more complete and concise 
review published in Ref.~\cite{concise}.
For a detailed introduction into the field and for an overview and 
definition of basic concepts, equations and 
references, the reader is therefore referred to~\cite{concise}.
Based on those results and on the current update,
a new world average of $\amz$, the value of the running 
(i.e. energy dependent) coupling constant at the rest energy of the $\z0$ 
boson, $\mz = 91.2$~GeV, will be given.

\section{NEW RESULTS}

New or updated measurements of $\as$ are available from almost all major 
classes of high energy particle reactions. 
In the following subsections, the respective results will be shortly reviewed.
In the summary of all relevant measurements of 
$\as$  given in Table~1, these updates and new results are underlined for
better visibility.

\subsection{Deep Inelastic Scattering (DIS)}

A new global analysis of all available precision data for deep inelastic
and related hard scattering processes includes recent measurements of
structure functions at HERA and of the inclusive jet cross sections at the
Tevatron \cite{martin2001}.
Apart from an improved determination of the gluon distribution function,
$\amz$ is quantified, in next-to-leading order (NLO) of perturbative QCD,
to be
$$\amz = 0.119 \pm 0.002\ (exp) \pm 0.003\ (theo)\ \ ,$$
where the theoretical error is estimated from alternative theoretical 
treatments, like NNLO fits or those including resummation techniques.

Other studies \cite{forte2002,kotikov2002}
based on subsets of data used in \cite{martin2001} provide results which 
are less precise but are compatible with the one above.

Several new QCD studies \cite{yndurain2001,kataev2001,maxwell2002}
based on data of the structure function 
$xF_3$ from neutrino-nucleon scattering \cite{ccfr97} were published, 
which are all based on complete next-next-to-leading order (NNLO) QCD
calculations. 
They result in: \newline \newline \noindent
%\begin{eqnarray} 
$\amz = 0.119 \pm 0.005\ (exp) ^{+0.005}_{-0.003}\ (theo) 
\cite{kataev2001}$ \newline \noindent
$\amz = 0.1153 \pm 0.0040\ (stat) \pm 0.0061\ (sys) \cite{yndurain2001} $
\newline \noindent
$\amz = 0.1196^{+0.0027}_{-0.0031}\ (exp)\ \cite{maxwell2002}. $ \newline
%\end{eqnarray}

The result from \cite{kataev2001} is based on Jacobian polynomials and 
moments of $xF_3$.
It includes renormalisation scale uncertainties in the 
assigned theoretical error.
The value of $\as$ obtained in \cite{yndurain2001} is based on a method 
using Berstein polynomials; it is systematically 
smaller and only compatible with \cite{kataev2001} within the assigned total
errors.
The most recent result \cite{maxwell2002} agrees with \cite{kataev2001}, 
however has no theoretical error assigned yet.
Until further clarification, the result of \cite{kataev2001} is taken 
as the final value of $\as$ from $xF_3$.

A previous study based on Bernstein polynomials and moments of the
structure function $F_2$ from deep inelastic electron- and muon-scattering 
data was recently updated \cite{yndurain2001}, giving
$\amz = 0.1166 \pm 0.009\ (stat) \pm 0.0010\ (sys)$
in NNLO QCD. 
The systematic error includes higher twist effects and an estimate of the 
NNNLO corrections, however does not include studies of renormalisation 
scale nor scheme uncertainties which cannot be estimated at the current 
state of the theoretical calculations \cite{yndurain-private}.
In order to account for the incomplete assessment of the theoretical 
uncertainty, the quoted value of $\pm 0.0010$ will be doubled such that
$$\amz = 0.1166 \pm 0.009\ (stat) \pm 0.0020\ (sys)$$
will be taken as the final result.

Finally, new results on $\as$ from jet production in deep ineleastic 
electron- (and/or positron-) proton scattering at HERA were reported
\cite{h1-jet-2001,zeus-jet-2002} which are combined to give, in NLO QCD,
$$\amz = 0.120 \pm 0.002\ (exp) \pm 0.004\ (theo)\ .$$

\subsection{$\epem$ Annihilation}

New results from studies of event shape distributions and jet production 
in $\epem$ annihilations where reported from a reanalysis of JADE data, at 
c.m. energies of 14 and of 22 GeV \cite{fernandez-2002}, in resummed NLO QCD, 
giving \newline \newline \noindent
$\as (14~GeV) = 0.170 \pm 0.005\ (exp) ^{+0.020}_{-0.016}\ (theo)$ 
\newline \noindent
$\as (22~GeV) = 0.151 \pm 0.004\ (exp) ^{+0.014}_{-0.012}\ (theo)\ .$
\newline \noindent

The four LEP experiments have reportet new results on $\as$ from event 
shapes and jet rates \cite{a-2002,d-2002,l-2002,o-2002}, in resummed NLO QCD,
in the c.m. energy range from 192 to 208~GeV.
Most of these results are still preliminary; here they are summarised as
\newline \newline \noindent
$\as (195~GeV) = 0.109 \pm 0.001\ (exp) \pm 0.005\ (theo)$
\newline \noindent
$\as (201~GeV) = 0.110 \pm 0.001\ (exp) \pm 0.005\ (theo)$
\newline \noindent
$\as (206~GeV) = 0.110 \pm 0.001\ (exp) \pm 0.005\ (theo).$
\newline \noindent

A dedicated QCD working group at LEP (LEPQCDWG) currently prepares 
combined values of $\as$ which are summarised over all four LEP 
experiments taking full account of correlations between observables, 
energies and 
experiments, using identical analysis methods and definitions of 
theoretical uncertainties, see e.g. \cite{lepqcd}.
Once the outcome of those studies will become official, they will superseed the 
respective numbers given in this and the previous \cite{concise} report.

The latest update of the combined LEP measurement of the ratio of the 
hadronic and the leptonic partial decay widths of the $\z0$ boson,
$R_l = \Gamma_{had} / \Gamma_{lept}$ results in
$$R_l = 20.767 \pm 0.025\ ,$$
from which - assuming strict validity of the electroweak standard model 
predictions for a top-quark mass of $M_t = 174.3 \pm 5.1~GeV$ 
and a Higgs boson mass of $115^{+100}_{-0}~GeV$ - $\as$ is 
being determined to be, in complete NNLO QCD, 
$$\amz = 0.1227 \pm 0.0038\ (exp) ^{+0.0029}_{-0.0005}\ (sys)\ ,$$
whereby the systematic uncertainty is determined as described in section 4.3
and Table~2 of \cite{concise}.

A novel determination of $\as$ from 4-jet observables \cite{a-4j}
in $\oaaa$ (i.e. NLO) QCD results in
$\amz = 0.1170 \pm 0.0001\ (stat) \pm 0.0013\ (sys)$, where the systematic 
error includes variation of the renormalisation scale factor between 0.5 
and 2.0, differences between two major hadronisation models, experimental 
systematic uncertainties and quark mass dependences.
At the claimed precision, the amount of systematic studies, however, seems 
rather optimistic; e.g. the hadronisation uncertainty may be accidentally 
small for the two (optimised) models being investigated.
The influence of model parameter variations should also have been included,
and higher order QCD uncertainties should be studied more intensively then 
by small variations of the renormalisation scale.
Until more experience with this type of analysis is available, the claimed 
systematic uncertainty is doubled such that the final result included 
in this summary is 
$$\amz = 0.1170 \pm 0.0001\ (stat) \pm 0.0026\ (sys)\ .$$

Finally, a new determination of $\as$ from measurements of the photon 
structure function $F_2^{\gamma}$ in two photon reactions at LEP
\cite{lep-2gamma} became available, where $\as$ was determined to be, in 
NLO QCD, 
$$\amz = 0.1198 \pm 0.0028\ (exp) ^{+0.0034}_{-0.0046}\ (theo).$$

%%%%%%%%%%%%%%%%%%%%%%%%%%%%%%%%%%%%%% Figure 1
\begin{figure}[htb]
%\vskip-20mm
\begin{center}
\epsfxsize7.5cm
\epsffile{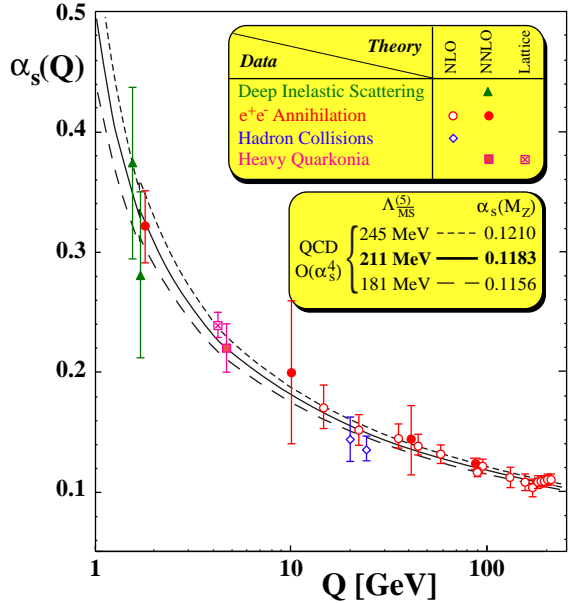}
\end{center}
\vskip-10mm
\caption{\label{asq}
Summary of measurements of $\asq$. 
Results which are based on fits of $\amz$ to data in $ranges$ of $Q$, 
assuming the QCD running of $\as$,
are not shown here but are included
in the overall summary of $\amz$, see Fig. 2 and Tab. 1
%\protect\cite{firstjets}
.}
\end{figure}
%%%%%%%%%%%%%%%%%%%%%%%%%%%%%%%%%%%%%%%% end Figure 1

\subsection{Hadron Collisions}

An updated study of inclusive jet production at the Tevatron \cite{cdf-jet}, 
in the transverse energy range of $40 < E_T < 250~GeV$ and based on NLO 
QCD, resultet in 
$$\amz = 0.118 ^{+0.008}_{-0.010}\ (exp) ^{+0.009}_{-0.008}\ (theo)\ ,$$
where the theoretical error includes renormalisation and parton 
density function uncertainties.

\subsection{Heavy Quarkonia Mass Splittings}

A new determination of $\as$ from $\Upsilon$ mass splittings and lattice 
QCD calculations with 3 dynamic quark flavours \cite{lgt-3}, 
which for the first time do not 
have to rely on extrapolations of calculations with 0 or 2 to 3 physical 
quark flavours, results in 
$$\amz = 0.121 \pm 0.000\ (exp) \pm 0.003\ (theo)\ .$$

%%%%%%%%%%%%%%%%%%%%%%%%%%%%%%%%%%%%%% Figure 2
\begin{figure}[tb]
\begin{center}
\epsfxsize7.0cm
\epsffile{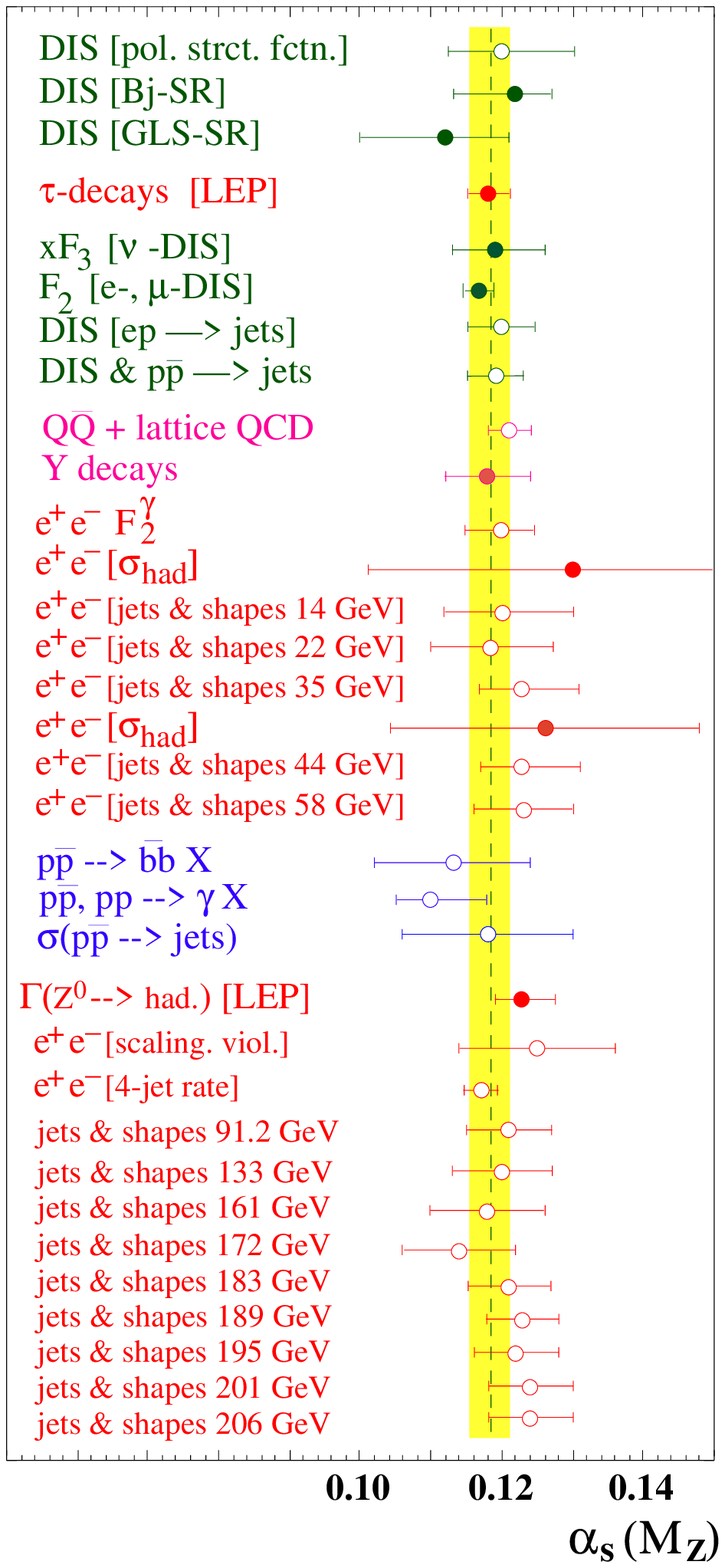}
\end{center}
\vskip-10mm
\caption{\label{asmz-2002}
Summary of measurements of $\amz$. 
Filled symbols represent results based on complete NNLO QCD calculations
%\protect\cite{firstjets}
.}
\end{figure}
%%%%%%%%%%%%%%%%%%%%%%%%%%%%%%%%%%%%%%%% end Figure 2

\section{SUMMARY AND THE NEW WORLD AVERAGE}

A summary of all measurements of $\as$, 
as discussed in \cite{concise} but 
with all updates and new measurements presented in the previous section, 
is given in Table~1.
The values of $\as (Q)$ are presented in Figure~1, as a 
function of the energy scale $Q$ where the measurement was carried 
out\footnote{Results which were determined from data in 
$ranges$ of energy $Q$ depend on the explicit assumption of 
the energy dependence of
$\as$ as predicted by QCD; they are not shown in Figure~1
but are included in Table 1 and in Figure 2.}.
As already seen in previous summaries of $\as$, the data provide 
significant evidence for the running of $\as$, in good agreement with the 
QCD prediction.

Therefore it is appropriate to extrapolate all results of $\as (Q)$ to a common 
value of energy, which is usually the rest energy (or mass, using the 
convention $c = 1$) of the $\z0$ boson,  $\mz$.
As done and described in \cite{concise}, the QCD evolution of $\as$ with 
energy, using the full 4-loop expression \cite{4-loop} with 3-loop matching 
\cite{matching} at the pole masses of the charm- and the bottom-quark,
$M_c = 1.7\ GeV$ and $M_b = 4.7\ GeV$, is applied to all results of $\as (Q)$ 
which were obtained at energy scales $Q \ne \mz$.

The corresponding values of $\amz$ are tabulated in the $4^{th}$ column of 
Table~1; column 5 and 6 indicate the  contributions of the experimental and 
the theoretical unceratinties to the overall errors assigned to $\amz$.
All values of $\amz$ are graphically displayed in Figure~2.
Within their individual uncertainties, there is perfect agreement between 
all results. 
This justifies to evaluate an overall world average value, $\wamz$.
As discussed e.g. in \cite{concise}, however, the combination of all these 
results to an overall average, and even more so for the overall
uncertainty to be assigned to this average,
is not trivial due to the supposedly large 
but unknown correlations between invidual results, especially through 
common prejudices and biases within the theoretical calculations.

For combining all or subsets of the results summarised in Table~1 into 
average values of $\amz$, the same procedures as utilised  
in \cite{concise} are being used:
\begin{itemize}
\item
An error weighted average and an \oq optimized
correlation" error is calculated from the error covariance matrix, assuming
an overall correlation factor between the total errors of all measurements.
This factor is adjusted so that the overall $\chi^2$ equals one per degree
of freedom~\cite{schmelling}.
The resulting mean values, overall uncertainties and optimized
correlation factors are given in columns 3 to 5 of
Table~\ref{tab:aserr}, respectively.
\item
For illustrative purposes only, an overall error is calculated assuming that all
measurements are entirely uncorrelated and all quoted errors are gaussian.
The results are displayed in column 6.
\item
The simple, unweighted root mean squared of the mean values of all
measurements is calculated  and shown in column 7, labelled \oq simple
rms".
\item
Assuming that each result of $\amz$ has a rectangular-shaped rather than a
gaussian probability
distribution, 
the resulting weights (the inverse of the square of the total
error) are summed up in a histogram, and the resulting $rms$ of that distribution
is quoted as \oq rms box" \cite{qcd97}.
\end{itemize}

Averages $\wamz$ for all and for subsets of $\as$-results, together 
with the corresponding uncertainties $\dwas$ are summarised in 
Table~\ref{tab:aserr}.
As already discussed in \cite{concise}, the overall uncertainties decrease 
if the averaging process is restricted to those which accomplished a 
minimum precision, i.e. a total error of $\Delta \as \le 0.008$, while the
value of $\wamz$ is almost unaffected by such a restriction - c.f. rows~1 
and~2.

The second observation is that there is one single result, the one derived from 
moments of structure functions $F_2$ \cite{yndurain2001}, which 
influences the average and its uncertainty most significantly. 
Therefore, in Table~\ref{tab:aserr}, averages for all and for subsets of data 
are also given excluding the result from $F_2$ (indicated by \oq $- F_2$").

There is a sufficiently large number of redults which is based on complete 
NNLO QCD, such that $\wamz$ can be reliably calculated from this subset 
(see rows 5 to 8 of Table~\ref{tab:aserr}).
Due to the improved completenes of the perturbation series, these results 
are believed to be more reliable and better defined than all the others 
which are complete to (resummed) NLO.
The new world average of $\amz$ is finally quoted from those NNLO results
which have total errors less than 0.008, giving
\begin{equation}
\wamz = 0.1183 \pm 0.0027\ ,
\end{equation}
c.f. row~6 of Table~\ref{tab:aserr}.
The average value is practically indentical to the final result of the 
previous summary \cite{concise}, $\wamz = 0.1184 \pm 0.0031$,
in spite of the many updates and new results presented above, while the 
overall assigned, correlated uncertainty decreased by about 10\%.

The result of $\wamz = 0.1183 \pm 0.0027$ corresponds to the following 
values of the QCD scale $\lamsb$ for different numbers of 
quark flavours $N_f$, 
evaluated using the full 4-loop expansion
of $\as$ and 3-loop matching at the quark thresholds (c.f. Equations~4 
and~8 of \cite{concise}):
\begin{eqnarray}
\lamsb^{N_f=5} &=& 211^{+34}_{-30}\ {\rm MeV} \nonumber \\
\lamsb^{N_f=4} &=& 294^{+42}_{-38}\ {\rm MeV} \nonumber \\
\lamsb^{N_f=3} &=& 336^{+42}_{-38}\ {\rm MeV} \ . \nonumber 
\end{eqnarray}

\section{CONCLUDING REMARKS}

There is a remarkable agreement of all single results with the new world 
average value; no result deviates by more than one standard deviation of 
its assigned uncertainty. 
This indicates that the uncertainties must be either overestimated 
or strongly correlated.
Because most quoted errors are dominated by theoretical uncertainties
which are based on estimates that are for sure non-gaussian, 
the latter of these two possibilities is probably true.

This taken for granted, the results should not be averaged using standard
methods of gaussian error propagation and error reduction, as done in other
reviews of $\as$ measurements \cite{pdg,hinchliffe}.  Such a procedure
systematically underestimates the resulting overall error on $\wamz$, even
if it is generously rounded up as done in
\cite{pdg,hinchliffe}.
The method of introducing an overall correlation between all results, such 
that the error matrix calculation provides a total $\chi^2$ per degree of 
freedom of unity \cite{schmelling}, as it is used to determine the results 
above, should provide - in the absence of any concrete knowledge of these 
correlations - a more realistic estimate of the overall remaining 
uncertainty $\dwas$.

Some observations made in the process of summarising and averaging are 
worth mentioning:

Both $\wamz$ and $\dwas$ increase by about 1\% of $\wamz$ if the result from 
moments of $F_2$ \cite{yndurain2001} 
is not included in the average, c.f. rows 1 and 3, rows
5 and 7 or rows 6 and 8 of Table~\ref{tab:aserr}.
This is the largest systematic change due to the omission of one 
single result which was observed in this study,
which is, however, fully compatible within the quoted errors and 
uncertainties.
Nevertheless, since the result from $F2$ lacks studies of renormalisation 
scale and scheme uncertainties and their inclusion in the quoted overall error, 
and since it's influence on the final average values is dominant - although 
the originally quoted theoretical uncertainty was doubled in this report - 
there is a grain of salt left when including this measurement in 
the overall summary. 
While the $F_2$ result $is$ included in the final value of $\wamz = 0.1183$,
it is worth mentioning that omission of that result leads to sligthly 
higher values of $\wamz \sim 0.1195$, c.f. Table~\ref{tab:aserr}. 

With very few exceptions, the most recent results summarised in this report 
(c.f. the underlined entries in Table~1) all tend towards values of $\amz 
\sim 0.120$, especially those from DIS processes and the one based 
on improved lattice gauge theory. 
In the past, those results typically arrived at values smaller than found 
e.g. in $\epem$ annihilation processes.
Accordingly, there is no significant systematic difference between 
results form DIS and from $\epem$ annihilations any more, c.f. rows 12 to 
13 of Table~\ref{tab:aserr}. 
The same is true for $\wamz$ determined from results based on complete NNLO 
or from (resummed) NLO QCD calulcations only, c.f. rows 5 and 9 or 6 and 10 of
Table~\ref{tab:aserr}.
These observations are even more exact when the result from moments of 
$F_2$ is not included, c.f. rows 12 and 13 as well as rows 7 and 9.

Concluding, and leaving aside the small effects discussed in this section,
the observed stability of $\wamz$ and the significant 
energy dependence of $\as$, which 
is in perfect agreement with the QCD prediction of the running of $\as$, 
constitute a major success of uncovering nature's intrinsic
basic parameters and features!

%%%%%%%%%%%%%%%%%%%%%%%%%%%%%%%%%%% Begin Table astab
\renewcommand{\arraystretch}{1.2}
\begin{table*}[htb]
{%\tiny
\caption{
World summary of measurements of $\as$ (status of September 2002):
DIS = deep inelastic scattering; GLS-SR = Gross-Llewellyn-Smith sum rule;
Bj-SR = Bjorken sum rule;
(N)NLO = (next-to-)next-to-leading order perturbation theory;
LGT = lattice gauge theory;
resum. = resummed NLO). 
New results and updates since the summary of 2000 \protect\cite{concise}
are underlined.
\label{tab:astab}}
\begin{center}
\begin{tabular}{|l|c|c|c|c c|c|}
   \hline 
  & Q & & &  \multicolumn{2}{c|}
{$\Delta \amz $} &  \\ %\cline{5-6}
Process & [GeV] & $\alpha_s(Q)$ &
  $ \amz$ & exp. & theor. & Theory \\
\hline \hline %\normalsize
DIS [pol. strct. fctn.] & 0.7 - 8 & & $0.120\ ^{+\ 0.010}
  _{-\ 0.008}$ & $^{+0.004}_{-0.005}$ & $^{+0.009}_{-0.006}$ & NLO \\
DIS [Bj-SR] & 1.58
  & $0.375\ ^{+\ 0.062}_{-\ 0.081}$ & $0.121\ ^{+\ 0.005}_{-\ 0.009}$ & 
  -- & -- & NNLO \\
DIS [GLS-SR] & 1.73
  & $0.280\ ^{+\ 0.070}_{-\ 0.068}$ & $0.112\ ^{+\ 0.009}_{-\ 0.012}$ & 
  $^{+0.008}_{-0.010}$ & $0.005$ & NNLO \\
$\tau$-decays 
  & 1.78 & $0.323 \pm 0.030$ & $0.1181 \pm 0.0031$
  & 0.0007 &  0.0030 & NNLO \\
\underline{DIS [$\nu$; ${\rm x F_3}$]}  & 2.2 - 12.3
  & 
   & $0.119\ ^{+\ 0.007}_{-\ 0.006}$   &
    $ 0.005 $ & $^{+0.005}_{-0.003}$ & NNLO \\
\underline{DIS [e/$\mu$; ${\rm F_2}$]}
     & 1.9 - 15.2 &      & $0.1166 \pm 0.0022$ & $ 0.0009$ &
     $ 0.0020$ & NNLO \\
\underline{DIS [e-p $\rightarrow$ jets]}
     & 6 - 100 &  & $0.120 \pm 0.005$ & $ 0.002$ &
     $0.004 $ & NLO \\
\underline{DIS \& $ \rm p\bar{p}\rightarrow$jets}
     & 1 - 400 &  & $0.119 \pm 0.004$ & $ 0.002$ &
     $0.003 $ & NLO \\
\underline{${\rm Q\overline{Q}}$ states}
     & 4.1 & $0.239\ ^{+\ 0.012}_{-\ 0.010}$ & $0.121 \pm 0.003 
     $ & 0.000 & 0.003
     & LGT \\
$\Upsilon$ decays
     & 4.75 & $0.217 \pm 0.021$ & $0.118 \pm 0.006
     $ & -- & -- & NNLO \\
\underline{$\epem$ [${\rm F^{\gamma}_2}$]}
     & 1.4 - 28 &  & $0.1198\ ^{+\ 0.0044}_{-\ 0.0054}$ 
     & 0.0028 & $^{+\ 0.0034}_{-\ 0.0046}$ & NLO \\
$\epem$ [$\sigma_{\rm had}$] 
     & 10.52 & $0.20\ \pm 0.06 $ & $0.130\ ^{+\ 0.021\ }_{-\ 0.029\ }$
     & $\ ^{+\ 0.021\ }_{-\ 0.029\ }$ & 0.002 & NNLO \\
\underline{$\epem$ [jets \& shapes]}
   & 14.0 & $0.170\ ^{+\ 0.021}_{-\ 0.017}$ &
   $0.120\ ^{+\ 0.010}_{-\ 0.008}$ &  0.002 & $^{+0.009}_{-0.008}$
   & resum \\
\underline{$\epem$ [jets \& shapes]}
   & 22.0 & $0.151\ ^{+\ 0.015}_{-\ 0.013}$ &
   $0.118\ ^{+\ 0.009}_{-\ 0.008}$ &  0.003 & $^{+0.009}_{-0.007}$
   & resum \\
$\epem$ [jets \& shapes] & 35.0 & $ 0.145\ ^{+\ 0.012}_{-\ 0.007}$ &
   $0.123\ ^{+\ 0.008}_{-\ 0.006}$ &  0.002 & $^{+0.008}_{-0.005}$
   & resum \\
$\epem$ [$\sigma_{\rm had}$]  & 42.4 &
 $0.144 \pm 0.029$ &
   $0.126 \pm 0.022$ & $0.022
   $ & 0.002 & NNLO \\
$\epem$ [jets \& shapes] & 44.0 & $ 0.139\ ^{+\ 0.011}_{-\ 0.008}$ &
   $0.123\ ^{+\ 0.008}_{-\ 0.006}$ & 0.003 & $^{+0.007}_{-0.005}$
   & resum \\
$\epem$ [jets \& shapes]  & 58.0 & $0.132\pm 0.008$ &
   $0.123 \pm 0.007$ & 0.003 & 0.007 & resum \\
%& & & & & & & \\
$\p\bar{\p} \rightarrow {\rm b\bar{b}X}$
    & 20.0 & $0.145\ ^{+\ 0.018\ }_{-\ 0.019\ }$ & $0.113 \pm 0.011$ 
    & $^{+\ 0.007}_{-\ 0.006}$ & $^{+\ 0.008}_{-\ 0.009}$ & NLO \\
${\rm p\bar{p},\ pp \rightarrow \gamma X}$  & 24.3 & $0.135
 \ ^{+\ 0.012}_{-\ 0.008}$ &
  $0.110\ ^{+\ 0.008\ }_{-\ 0.005\ }$ & 0.004 &
  $^{+\ 0.007}_{-\ 0.003}$ & NLO \\
%${\rm p\bar{p} \rightarrow W\ jets}$  & 80.6 & $0.123 \pm 0.025$ &
%  $0.121\pm 0.024$ & 0.017 & 0.016 & NLO \\
\underline{${\sigma (\rm p\bar{p} \rightarrow\  jets)}$}
  & 40 - 250 &  &
  $0.118\pm 0.012$ & $^{+\ 0.008}_{-\ 0.010}$ & $^{+\ 0.009}_{-\ 0.008}$ & NLO \\
\underline{$\epem$ [$\Gamma (\z0 \rightarrow {\rm had.})$]}
    & 91.2 & $0.1227^{+\ 0.0048}_{-\ 0.0038}$ & 
    $0.1227^{+\ 0.0048}_{-\ 0.0038}$ &
   $ 0.0038$ & $^{+0.0029}_{-0.0005}$ & NNLO \\
$\epem$ scal. viol. & 14 - 91.2 &  & $0.125 \pm 0.011$ & 
   $^{+\ 0.006}_{-\ 0.007}$ & 0.009 & NLO \\
\underline{$\epem$ 4-jet rate}
  &  91.2 & $0.1170 \pm 0.0026$ & $0.1170 \pm 0.0026$ & 
  0.0001 & 0.0026 & NLO \\
$\epem$ [jets \& shapes] &
    91.2 & $0.121 \pm 0.006$ & $0.121 \pm 0.006$ & $ 0.001$ & $
0.006$ & resum \\
%& & & & & & & \\
$\epem$ [jets \& shapes]  & 133 & $0.113\pm 0.008$ &
   $0.120 \pm 0.007$ & 0.003 & 0.006 & resum \\
$\epem$ [jets \& shapes]  & 161 & $0.109\pm 0.007$ &
   $0.118 \pm 0.008$ & 0.005 & 0.006 & resum \\
$\epem$ [jets \& shapes]  & 172 & $0.104\pm 0.007$ &
   $0.114 \pm 0.008$ & 0.005 & 0.006 & resum \\
$\epem$ [jets \& shapes]  & 183 & $0.109\pm 0.005$ &
   $0.121 \pm 0.006$ & 0.002 & 0.005 & resum \\
$\epem$ [jets \& shapes] & 189 & $0.109\pm 0.004$ &
   $0.121 \pm 0.005$ & 0.001 & 0.005 & resum \\
\underline{$\epem$ [jets \& shapes]}
   & 195 & $0.109\pm 0.005$ &
   $ 0.122\pm 0.006$ & 0.001 & 0.006 & resum \\
\underline{$\epem$ [jets \& shapes]}
   & 201 & $0.110\pm 0.005$ &
   $ 0.124\pm 0.006$ & 0.002 & 0.006 & resum \\
\underline{$\epem$ [jets \& shapes]}
   & 206 & $0.110\pm 0.005$ &
   $ 0.124\pm 0.006$ & 0.001 & 0.006 & resum \\
& & & & & & \\
\hline
\end{tabular}
\end{center}
}
\end{table*}
%%%%%%%%%%%%%%%%%%%%%%%%%%%%%%%%%%%%%% end table astab

%%%%%%%%%%%%%%%%%%%%%%%%%%%%%%%%%%%%%% start table aserr
\renewcommand{\arraystretch}{1.3}
\begin{table*}[htb]
\caption{
Average values of $\wamz$ and averaged uncertainties, for several
methods to estimate the latter, and for several subsamples 
of the available data. 
The result printed in bold-face is taken as the new
world average value of $\wamz$.\label{tab:aserr} }
\begin{center}
  {%\tiny 
\begin{tabular}{|c|l|c|c|c||c|c|c|}
   \hline
& & & opt. corr. & overall & uncorrel. & simple rms &  rms box \\
row & sample \hfill (entries)& $\wamz$ & $\dwas$ & correl. &
  $\dwas$ & $\dwas$ 
  & $\dwas$ \\
\hline
1 & all \hfill (33)     & 0.1189 & 0.0037 & 0.65 & 0.0009 & 0.0042 & 0.0051\\
2 &\ "\ $\Delta \as\le 0.008$ \hfill (25)
                        & 0.1189 & 0.0033 & 0.58 & 0.0009 & 0.0035 & 0.0044\\
3 & all - $F_2$ \hfill (32)
                        & 0.1194 & 0.0042 & 0.67 & 0.0010 & 0.0042 & 0.0052\\
4 &\ "\ $\Delta \as\le 0.008$\hfill (24)
                        & 0.1194 & 0.0037 & 0.62 & 0.0010 & 0.0035 & 0.0045\\
 & & & & & & &\\
5 & NNLO only \hfill (9) & 0.1183 & 0.0031 & 0.67 & 0.0015 & 0.0053 & 0.0049\\
6 &\ "\ $\Delta \as\le 0.008$ \hfill (6)
                        &\bf  0.1183 & \bf 0.0027 & 0.60 
                        & 0.0015 & 0.0022 & 0.0035\\
7 & NNLO - $F_2$ \hfill (8)
                        & 0.1196 & 0.0042 & 0.75 & 0.0020 & 0.0055 & 0.0052\\
8 &\ "\ $\Delta \as\le 0.008$ \hfill (5)
                        & 0.1197 & 0.0038 & 0.75 & 0.0020 & 0.0020 & 0.0037\\
&  & & & & & &\\
9 & NLO only \hfill (24)
                        & 0.1194 & 0.0042 & 0.64 & 0.0011 & 0.0038 & 0.0050\\
10 &\ "\ $\Delta \as\le 0.008$\hfill (19)
                        & 0.1193 & 0.0037 & 0.57 & 0.0012 & 0.0038 & 0.0047\\

&  & & & & & &\\
11 & DIS only \hfill (7) & 0.1178 & 0.0034 & 0.81 & 0.0016 & 0.0031 & 0.0042\\
12 & DIS - $F_2$\hfill (6) & 0.1192 & 0.0054 & 0.90 & 0.0024 & 0.0033 & 0.0049\\
13 & $\epem$ only\hfill (22) & 0.1195 & 0.0041 & 0.66 & 0.0012 & 0.0040 & 0.0050\\
% & & & & & & &\\
\hline
\end{tabular} }
\end{center} 
\end{table*}
%%%%%%%%%%%%%%%%%%%%%%%%%%%%%%%%%%%%%%%%%%%% end table aserr

\end{document}